\documentclass[a4paper,11pt]{article}
\usepackage{pos}

\usepackage{color}

\title{The $B_c$ decay rate in the Standard Model}
\author*[a]{Jason~Aebischer}
\author[b]{Benjam\'in Grinstein}

\affiliation[a]{Physik-Institut, Universit\"at Z\"urich, CH-8057 Z\"urich, Switzerland}

\affiliation[b]{Department of Physics, University of California at San Diego,
    La Jolla, CA 92093, USA}

\emailAdd{jason.aebischer@physik.uzh.ch}
\emailAdd{bgrinstein@ucsd.edu}

\abstract{The Standard Model decay rate of the $B_c$ meson is discussed together with a novel approach based on the usage of experimental data in combination with an operator product expansion. In the new method differences of $B,\,D$ and $B_c$ meson decay rates are considered, for which the free-quark contributions drop out, leading to a reduction of the theory prediction.}

\FullConference{%
  Presented at the 30th International Symposium on Lepton Photon Interactions at High Energies, hosted by the University of Manchester, 10-14 January 2022.
}

\begin{document}
\maketitle

\section{Introduction}

\noindent
The decay of the $B_c=(\overline b c)$ meson, made up of two different heavy quarks, is described using Non-Relativistic QCD (NRQCD), where an expansion around the small velocities of the two non-relativistic quarks is performed. After matching the relevant QCD operators onto the NRQCD Lagrangian by integrating out the anti-particles of the respective quarks, this approach allows to perform a systematic velocity expansion of the resulting NRQCD operators, which can be carried out up to any given order and where the truncation uncertainty can be estimated. Furthermore, the symmetry properties of NRQCD allow to relate the relevant matrix elements of the four-quark operators to two parameters.

\noindent
In combination with the Optical Theorem and an Operator Product expansion (OPE) performed on the forward scattering matrix element of the $B_c$ meson, this framework allows to predict the $B_c$ decay rate in a consistent manner. This OPE \cite{Beneke:1996xe,Bigi:1995fs,Chang:2000ac} approach leads to similar results as those obtained using QCD Sum Rules \cite{Kiselev:2000pp} or Potential models \cite{Gershtein:1994jw}, which are all in the ball park of the experimental value of the decay rate.
Experimentally, the $B_c$ decay rate has been determined with rather small uncertainties by the LHCb \cite{LHCb:2014ilr,LHCb:2014glo} and CMS \cite{CMS:2017ygm} collaborations and averages to a value of

\begin{equation}\label{eq:Gexp}
  \Gamma_{B_c}^\text{exp} = 1.961(35) \,\text{ps}^{-1}\,.
\end{equation}

\noindent
The $B_c$ decay rate is an interesting case to study, since it allows to put stringent bounds on New Physics models in the context of the $R(D)$ and $R(D^*)$ anomalies. Examples include scalar Leptoquarks and Two-Higgs-Doublet models \cite{Alonso:2016oyd,Blanke:2018yud,Blanke:2019qrx}.

\noindent
In the following we will summarize the results from an updated OPE computation of the $B_c$ decay rate \cite{Aebischer:2021ilm,Aebischer:2022fle}. Finally we outline a new method to obtain $\Gamma_{B_c}$, using experimental results as well as non-perturbative expansions of the $B_c$, $B$ and $D$ mesons.

\section{Results}

\noindent
Following the discussion in \cite{Aebischer:2021ilm}, we summarize the obtained values of the $B_c$ decay rate together with the uncertainties in three different mass schemes, namely the $\overline{\text{MS}}$, the meson and the Upsilon scheme. Neglecting the strange-quark mass one finds:

\begin{equation}\label{eq:ms0}
\begin{aligned}
  \Gamma^{\overline{\text{MS}}}_{B_c} &= (1.58\pm 0.40|^{\mu}\pm
  0.08|^{\text{n.p.}}\pm 0.02|^{\overline{m}}\pm 0.01|^{V_{cb}})\,\,\text{ps}^{-1}\,, \\
  \Gamma^{\text{meson}}_{B_c} &= ( 1.77\pm 0.25|^{\mu}\pm 0.20|^{\text{n.p.}} \pm 0.01|^{V_{cb}})\,\,\text{ps}^{-1} \,, \\
  \Gamma^{\text{Upsilon}}_{B_c} &= (2.51\pm 0.19|^{\mu}\pm 0.21|^{\text{n.p.}}\pm 0.01|^{V_{cb}})\,\,\text{ps}^{-1} \,.
\end{aligned}
\end{equation}

\noindent
Here the main uncertainties result from the scale dependence, indicated by $\mu$ in the above equation. It can be reduced by including higher-order QCD corrections to the free-quark decay rates, as well as to the Wilson coefficients involved. The second largest uncertainty, indicated by $n.p.$ in eq.~\eqref{eq:ms0}, stems from neglected higher-order corrections in the NRQCD expansion as well as from uncertainties of the non-perturbative parameter. In the $\overline{\text{MS}}$ scheme a non-negligible part of the uncertainty results from the $\overline{\text{MS}}$ masses of the $\overline b$- and $c$ quarks. Finally, there is an uncertainty due to the input parameters, which is dominated by $V_{cb}$.

\noindent
When including the strange-quark mass in the calculation for the free $c$-quark decays, the central values of the decay rate is reduced by about 7\%. We obtain in the three different schemes

\begin{equation}\label{eq:msnon0}
\begin{aligned}
  \Gamma^{\overline{\text{MS}}}_{B_c} &= (1.51\pm 0.38|^{\mu}\pm 0.08|^{\text{n.p.}}\pm 0.02|^{\overline{m}}  \pm0.01|^{m_s}\pm 0.01|^{V_{cb}})\,\,\text{ps}^{-1}\,, \\
  \Gamma^{\text{meson}}_{B_c} &= (1.70\pm 0.24|^{\mu}\pm 0.20|^{\text{n.p.}} \pm0.01|^{m_s}\pm 0.01|^{V_{cb}})\,\,\text{ps}^{-1} \,,  \\
  \Gamma^{\text{Upsilon}}_{B_c} &= (2.40\pm 0.19|^{\mu}\pm 0.21|^{\text{n.p.}} \pm0.01|^{m_s}\pm 0.01|^{V_{cb}})\,\,\text{ps}^{-1} \,.
\end{aligned}
\end{equation}

\noindent
Besides the uncertainties mentioned above we have indicated the uncertainty due to $m_s$.

\noindent
The results in eq.~\eqref{eq:msnon0} are within the respective uncertainties consistent with each other and with the experimental value in eq.~\eqref{eq:Gexp}. There is however a rather wide spread among the three different mass schemes used. One strategy to improve on the precision of the theory result is to reduce the uncertainty due to scale-dependence. In the next section we will discuss a novel approach which follows this route.

\section{Novel determination of $\Gamma_{B_c}$}\label{sec:newmethod}

\noindent
As discussed in the previous section, the main theory uncertainty stems from the renormalization scale dependence. It results mainly from the free-quark decay rate, which is the leading term in the non-perturbative expansion of the decay rate of a meson $H_Q$ with heavy quark $Q$:

\begin{equation}\label{eq:GM}
  \Gamma(H_Q) = \Gamma_Q^{(0)}+\Gamma^{n.p.}(H_Q)+\Gamma^{\text{WA}+\text{PI}}(H_Q)+\mathcal{O}(\frac{1}{m_Q^4})\,,
\end{equation}

\noindent
where the second term includes non-perturbative corrections and the third term contains Weak Annihilation and Pauli Interference contributions. The expansion in eq.~\eqref{eq:GM} can be carried out not only for the $B_c$ meson, but also for the $B$ and $D$ mesons. Taking now the difference of the three different decay rates leads to:

\begin{align}\label{eq:diff}
  \Gamma(B)+\Gamma(D)-\Gamma(B_c) &= \Gamma^{n.p.}(B)+\Gamma^{n.p.}(D)-\Gamma^{n.p.}(B_c) \nonumber \\
  &+\,\Gamma^{\text{WA}+\text{PI}}(B)+\Gamma^{\text{WA}+\text{PI}}(D)-\Gamma^{\text{WA}+\text{PI}}(B_c)\,.
\end{align}

\noindent
Since the free quark decay rate is independent of the meson state, it drops out on the right-hand side of eq.~\eqref{eq:diff}, thereby reducing the uncertainty due to scale-dependence. For the computation of $\Gamma(B_c)$, the decay rates of the $B$ and $D$ mesons can be taken from experiment, whereas the right-hand side of eq.~\eqref{eq:diff} can be computed using non-perturbative methods. The computation can be carried out for charged or neutral $B$ and $D$ mesons, leading in principle to four different ways to compute $\Gamma(B_c)$. In Tab.~\ref{tab:res} we show the results for the $B_c$ decay rate in the meson scheme, obtained using the four different channels \cite{Aebischer:2021eio}.

\begin{table}[t]
\centering
 \begin{tabular}{|l |c |c |c |c|}
 \hline
 & $B^0,D^0$ & $B^+,D^0$ & $B^0,D^+$ & $B^+,D^+$ \\ [0.5ex]
 \hline \hline
$\Gamma^{\text{meson}}_{B_c}$ & 3.03 $\pm$ 0.51 $\text{ps}^{-1}$ & 3.03 $\pm$ 0.53 $\text{ps}^{-1}$ & 3.33 $\pm$ 1.29 $\text{ps}^{-1}$ & 3.33 $\pm$ 1.32 $\text{ps}^{-1}$ \\
 \hline
 \end{tabular}
 \caption{\small
Results obtained using the novel approach discussed in sec.~\ref{sec:newmethod} in the meson scheme, using four different combinations of $B$ and $D$ mesons.
}
  \label{tab:res}
\end{table}

\noindent
The results from this novel approach are in tension with the experimental result in eq.~\eqref{eq:Gexp}. Several reasons can be put forward to explain this disparity:

\begin{enumerate}
  \item The uncertainties from NLO corrections to Wilson coefficients and free quark decay rates might be underestimated
  \item Eye-graph contributions, neglected in lattice computations of matrix elements that we use~\cite{Becirevic:2001fy}, but estimated to be small using HQET sum rules \cite{King:2021jsq}
  \item Unexpectedly large contributions from higher dimension operators in the $1/m_Q$ expansion~\cite{King:2021xqp}
  \item Violation of quark-hadron duality
\end{enumerate}

\noindent
A thorough analysis of the above mentioned points is in order to determine the reason for the discrepancy between the results and experiment.

\section{Summary}\label{sec:summary}

\noindent
We have outlined the OPE approach to determine the $B_c$ decay rate in the Standard Model, together with the obtained results in three different mass schemes. The obtained results in the $\overline{\text{MS}}$, the meson and the Upsilon scheme are compatible with each other and with the experimental value. There is however a wide spread among the central values in the three different mass schemes, where the main uncertainties result from neglected NLO QCD corrections as well as non-perturbative corrections.

\noindent
Secondly we discussed a novel method to determine $\Gamma_{B_c}$. It is based on differences of $B,\,D$ and $B_c$ decay rates and allows to reduce the scale-dependence uncertainty. The disparity of the obtained results in this approach and the experimental decay rate might have several reasons, including underestimation of uncertainties, large eye-graph contributions, corrections from dimension-seven operators to the charm decays or quark-hadron duality violation.

\small
\bibliographystyle{JHEP}
\bibliography{AGbib}

\end{document}